\documentstyle[12pt]{article}
\begin{document}
\newcommand{\bea}{\begin{eqnarray}}
\newcommand{\eea}{\end{eqnarray}}
\renewcommand{\thefootnote}{\fnsymbol{footnote}}
\newcommand{\be}{\begin{equation}}      \newcommand{\ee}{\end{equation}}
\newcommand{\st}{\scriptsize}
\newcommand{\fs}{\footnotesize}
\vspace{1.3in}
\begin{center}
\baselineskip 0.3in

{\large \bf Spectroscopic investigation of spin zero homonuclear and heteronuclear 
molecules }

\vspace {0.3in}
{ Nagalakshmi A. Rao }

\vspace{-0.2in}
{\it Department of Physics, Government Science College,}

\vspace{-0.2in}
{\it Bangalore-560001,Karnataka, India.}

\vspace{-0.2in}
{drnarao@gmail.com}
 
\vspace{0.2in}
{ B. A. Kagali }

\vspace{-0.2in}
{\it Department of Physics, Jnanabharathi Campus, Bangalore University,}

\vspace{-0.2in}
{\it Bangalore-560056,Karnataka, India.}

\vspace{-0.2in}
{bakagali@hotmail.com}
\end{center}

\vspace{0.3in}
\hspace{2.3in}

In the present article, we introduce a model to investigate the energy spectrum of a 
relativistic rotor by considering the Klein-Gordon Hamiltonian. Rotational spectral lines
are a signature of homonuclear and heteronuclear systems and play a key role in understanding 
diatomic molecules. We show that the energy-correction term arising due to unequal masses
influences the line separation. Determining the rotational constant enables one to calculate the moment 
of inertia and bond length of the molecule.

\vspace{0.6in}

\vspace{0.1in}

\vspace{0.1in}
\noindent
{\bf Comments:} Latex, 11 Pages, No figures.

\vspace{0.6in}
\noindent

\setcounter{section}{0}
\newpage
\indent
{\section{\large \bf Introduction}}

	The rigid rotator is a well-known problem in classical mechanics$^{1}$ as also in quantum
mechanics.The quantum theory of the band spectra of diatomic molecules began  with the quantisation of angular momentum 
of the rigid rotator. Much of spectroscopic calculation relies on rigid rotator models. 
It is well-known that there exists a fundamental difference between  atomic spectra and  molecular spectra. While 
for atoms the line separation in a series decreases rapidly, for molecular spectra in the infrared, it 
is approximately a constant. Actually, for linear diatomic molecules, two additional modes of
 motion, namely rotation and vibration, not present in atoms, are possible. In literature, the energy spectrum of such a system is investigated 
non-relativistically. Presently, we propose to view the model in the relativistic framework. 

After the classic application of 
relativistic wave equations to the atomic spectrum of hydrogen atom, there were several other applications of the 
relativistic wave equations to well-known problems of non-relativistic quantum mechanics. In the simplest from, one can 
cite the work of Sauter$^{2}$, who treated the problem of a particle in an uniform field relativistically. Srinivasa Rao 
et.al.$^{3}$, have dealt with the relativistic rigid rotator in an elegant way, treating the rotor as a dumb bell with equal 
masses. Following the prescription of Rao, we extend the treatment to a linear rotator with unequal masses. The foregoing 
discussion is applicable to any heteronuclear system, special case of which is the homonuclear species. Such studies, 
apart from being pedagogical in nature are potentially exciting and shed light on the spectra of diatomic molecules. 

\section{\large \bf  Klein-Gordon Rotor}
\setcounter{equation}{0}
\indent

	In the simplest model of a rotator, we consider two atoms of masses $m_{1}$ and $m_{2}$ separated by distance $r$.
\indent

The Klein-Gordon Hamiltonian for a system of two non-interacting particles may be written as 
\vspace{-0.2in}
\bea
H=c\sqrt {m_{1}^{2}c^{2}+\vec p_{1}^{\>\,2}}+c\sqrt {m_{2}^{2}c^{2}+\vec
p_{2}^{\>\,2}}
\eea
where $\vec p_{1}$ and $\vec p_{2}$ refer to the momenta of the two particles.

\indent
Introducing the centre of mass co-ordinate $\vec R$ and the relative co-ordinate $\vec r$, the two body problem may be 
reduced to an equivalent one-body problem. The relative co-ordinate, $\vec r : (x,y,z)$ is expressed as 
\vspace{-0.2in}
\bea
\vec r=\vec r_{1}-\vec r_{2}
\eea
and the centre of mass co-ordinate $\vec R :(X,Y,Z)$ is given by 
\bea
\vec R={m_{1}\vec r_{1}\ +{\ m}_{2}\vec r_{2}\over m_{1}+m_{2}}.
\eea
\indent
It is straightforward to verify the following relations 
\bea
{\partial \over \partial x_{1}}={m_{1}\over M}\ {\partial \over \partial
X}+{\partial \over \partial x}
\eea
\bea
{\partial \over \partial x_{2}}={m_{2}\over M}\ {\partial \over \partial
X}-{\partial \over \partial x}
\eea
where $M=m_{1}+m_{2}$ is implied.

\indent
The eigenvalue equation, $H\psi =W\psi $ may be written as 
\bea
\left\{ c\sqrt {m_{1}^{2}c^{2}-\hbar ^{2}\nabla _{1}^{2}}+c\sqrt
{m_{2}^{2}c^{2}-\hbar ^{2}\nabla _{2}^{2}}-W\right\} \psi =0
\eea
using $\vec p_{1}=-i\hbar \vec \nabla _{1}$ and $\vec p_{2}=-i\hbar \vec \nabla _{2}$ in Eqn.(1). Operating on the left 
by $(H+W)$, we obtain, 
\vspace{-0.2in}
\bea
\left(W_{1}+{H}_{1}\right)\psi =0
\eea  
with \ \ \ $W_{1}=c^{2}\left(m_{1}^{2}c^{2}-\hbar ^{2}\nabla_{1}^{2}\right)+c^{2}\left(m_{2}^{2}c^{2}-\hbar ^{2}\nabla
_{2}^{2}\right)-W^{2}$

\noindent
and \ \ \ \ $H_{1}\,=2c^{2}\sqrt {m_{1}^{2}c^{2}-\hbar ^{2}\nabla _{1}^{2}}\sqrt {m_{2}^{2}c^{2}-\hbar ^{2}\nabla _{2}^{2}}.$

To eliminate the square root operators in $H_{1}$, we operate on the left of Eqn.(7) by $(W_{1}-H_{1})$ and obtain
\vspace{-0.2in}
\bea
\left(W_{1}^{2}-H_{1}^{2}\right)\psi =0.
\eea
\indent
The operator equation of the above is
\bea
\left[\!\left\{ {\!c}^{2}\!\left(m_{1}^{2}\!-\!\hbar ^{2}{\nabla
}_{1}^{2}\right)^{2}\!\!+\!c^{2}\!\left(m_{2}^{2}\!-\!\hbar ^{2}{\nabla
}_{2}^{2}\right)^{\!2}\!-\!{W}^{2}\!\right\}
^{\!2}\!-\!4c^{4}\!\left(m_{1}^{2}c^{2}\!-\!{\hbar }^{2}{\nabla
}_{1}^{2}\right)\!\!\left(m_{2}^{2}c^{2}\!-\!{\hbar }^{2}{\nabla
}_{2}^{2}\right)\!\right]\!\psi \!=\!0.
\eea
The above equation simplifies to 
$$\left[c^{4}\left(m_{1}^{2}c^{2}-\hbar ^{2}{\nabla
}_{1}^{2}\right)^{2}+c^{4}\left(m_{2}^{2}c^{2}-\hbar ^{2}{\nabla
}_{2}^{2}\right)^{2}+{W}^{4}-c^{4}\left(m_{1}^{2}c^{2}-{\hbar }^{2}{\nabla
}_{1}^{2}\right)\left(m_{2}^{2}c^{2}-{\hbar }^{2}{\nabla}_{2}^{2}\right)\right.$$
\bea
\ \ \ \ \ \ \ \ \ \ \ \ \ \ \ \ \ \ \ \left.-2{W}^{2}{c}^{2}\left(m_{1}^{2}c^{2}-{\hbar }^{2}{\nabla
}_{1}^{2}\right)-2W^{2}{c}^{2}\left(m_{2}^{2}{c}^{2}-{\hbar }^{2}{\nabla
}_{2}^{2}\right)\right]\psi =0.
\eea
Noting that $\vec \nabla _{1}=-\vec \nabla _{2}=\nabla $ implies that $\nabla _{1}^{2}=\nabla _{2}^{2}=\nabla ^{2}.$ 

Ignoring the translational motion of the system and 
holding the centre of mass at rest, Eqn.(10) reduces to 
\bea
\left[{W}^{4}\!+\!\!\left(m_{1}^{4}c^{8}+\!{m_{2}^{4}}c^{8}-\!2{m}_{1}^{2}{m}_{2}^{2}c^{8}\right)\!-\!2{W}^{2}m_{1}^{2}{c}^{4}\!-\!2W^{2}{m}_{2}^{2}{c}^{4}\!+\!4W^{2}{c}^{2}{\hbar
}^{2}{\nabla }^{2}\right]\!\psi \!=\!0.
\eea
It may well be checked that in the case of equal masses, $m_{1}=m_{2}=m_{0}$, the above equation reduces to 
\bea
\nabla ^{2}\psi\, +\,{\left(W^{2}-4m_{0}^{2}{c}^{4}\right)\over 4{c}^{2}{\hbar
}^{2}}\psi =0,  
\eea
which is in agreement with the work of Rao et.al$^{2}$.

\indent
The more general Eqn.(11) may be written in an elegant form as 
\bea
\left[\nabla
^{2}\>+\,\ \!{W^{2}-2\left(m_{1}^{2}c^{4}+m_{2}^{2}c^{4}\right)\over
4c^{2}{\hbar }^{2}}\,+\,{\left(m_{1}^{2}c^{4}-m_{2}^{2}c^{4}\right)^{2}\over
4{{W}^{2}c}^{2}{\hbar }^{2}}\right]\psi =0
\eea
\indent
To test the validity of this equation, we consider various cases.

\noindent
{\large \bf 2.1 Single Particle Rotor}

\indent
For a single particle of rest mass $m_{0}$, rotating about a fixed origin, Eqn.(6), with $m_{1}=m_{0}$ and $m_{2}=0,$ 
would yield,
\bea
\nabla ^{2}\psi +\left(W^{2}-m_{0}^{2}c^{4}\over c^{2}{\hbar }^{2}\right)\psi =0  
\eea
\indent
Expressing $\nabla^{2}$ in spherical polar co-ordinates and noting that {\Large ${\partial R\over \partial r}$}$=0$, we write
\bea
{1\over r^{2}{\rm sin}\theta }\ {\partial \over \partial \theta }\left({\rm sin}\theta
\ {\partial \psi \over \partial \theta }\right)+{1\over r^{2}{\rm sin}^{2}\theta
}\ {\partial ^{2}\psi \over \partial \phi ^{2}}+\left(W^{2}-{\epsilon }^{2}\over
{c}^{2}{\hbar }^{2}\right)\psi =0
\eea 
where $\epsilon =m_{0}c^{2}$.

With $r=a,$ $I=m_{0}a^{2}$, the moment of inertia of the particle about an axis passing through the origin, the above 
equation may be written as 
\bea
{1\over {\rm sin}\theta }\ {\partial \over \partial \theta }\left({\rm sin}\theta
\ {\partial \psi \over \partial \theta }\right)+{1\over {\rm sin}^{2}\theta
}\ {\partial ^{2}\psi \over \partial \phi ^{2}}+{I\left(W^{2}-{\epsilon
}^{2}\right)\over \epsilon {\hbar }^{2}}\ \psi =0
\eea
Comparing this with the corresponding non-relativistic Schrodinger equation, we immediately obtain,
\bea
{I\over \hbar ^{2}\epsilon }\left(W^{2}-\epsilon ^{2}\right)=l(l+1),
\eea
where $l$ is an integer.

\indent
It follows from above that the energy eigenvalues are                       
\bea
W= \epsilon \sqrt {1+{l\left(l+1\right)\hbar ^{2}\over I\epsilon }}
\eea
\noindent
{\large  \bf Non-relativistic limit :}

\indent
Taylor expansion of Eqn. (18) gives
\bea
W\approx \ \epsilon \left(1+{1\over 2}{l\left(l+1\right)\hbar ^{2}\over I\epsilon }\right)
\eea 
Defining $W-\epsilon \ =E_{Nr},$ we get the well-known expression for the energies
\bea
E_{Nr}={l\left(l+1\right)h^{2}\over 8\pi ^{2}I},\ where\ l=0,1,2,...
\eea
\noindent
{\large \bf 2.2 Homonuclear system }

\indent
For a system of two particles of equal masses rotating about the centre of mass, 

\vspace{-0.1in}
$m_{1}=m_{2}=m_{0}\ ,\ I=\mu a^{2}$ where the reduced mass $\mu =${\Large ${m_{o}\over 2}$} and $\epsilon=2m_{0}c^{2},$ the rest energy 
of the system, Eqn.(13) transforms into
\bea
{\nabla }^{2}\psi \,+\>{\left(W^{2}-\epsilon ^{2}\right)m\over 2\hbar ^{2}\epsilon }\psi=0
\eea
\noindent
which is the appropriate Klein-Gordon equation of the system. As before the eigenenergies correspond to Eqn.(18).

\noindent
{\large \bf 2.3 Heteronuclear System}

\indent
A more general case is a system of two non-interacting particles of unequal masses rotating about their centre of mass. The 
energy eigen-vlaues are obtained by solving Eqn.(13).

\indent
Expressing $\nabla^{2}$ in spherical polar coordinates, as before and observing that it transforms to the 
{\large $\left(-L^{2}\over \hbar ^{2}\right)$} operator with eigenvalues $l(l+1),$ we write 
\bea
{a^{2}\left\{ {W}^{2}-2\left(m_{1}^{2}{c}^{4}+m_{2}^{2}{c}^{4}\right)\right\}
\over 4{c}^{2}{\hbar
}^{2}} \ + \ {a^{2}\left(m_{1}^{2}{c}^{4}-m_{2}^{2}{c}^{4}\right)^{2}\over
4{W}^{2}{c}^{2}{\hbar }^{2}} \ = \ l(l+1)
\eea
\indent
Writing $\alpha=m_{1}c^{2} \ , \ \beta =m_{2}c^{2},$ it is obvious to check that the above equation may be written as 
\vspace{-0.2in}
\bea
a^{2}\ W^{4}-\left(A+B\right)W^{2}+C=0,
\eea
where
$$\matrix{A=2a^{2}\left(\alpha ^{2}+{\beta }^{2}\right)\cr \cr B=4l\left(l+1\right){c}^{2}{\hbar }^{2}\cr \cr 
C={a}^{2}\left(\alpha ^{2}-{\beta}^{2}\right)^{2}\cr } $$
\indent
 Eqn.(23) is solved using Mathematica$^{4}$ . Obviously, from the solution of the above equation
 for equal masses one should recover the energies of the homonuclear system. 
Hence of the four solutions, two of them being zero and the other being 
negative, are rejected and the only admissible solution is 
\bea
W\!\!=\!\!\!\left[\!\left(m_{\!1}^{\!2}c^{4}\!+\!m_{\!2}^{\!2}c^{4}\!\right)\!\!+\!{2l\over
a^{2}}\left(l\!+\!\!1\!\right)\!c^{2}{\hbar }^{\!2}\!+\!{2\over a^{2}}\!\sqrt
{\!\left(\!m_{\!1}^{\!2}c^{4}{a}^{\!2}\!+\!c^{2}{\hbar
}^{\!2}l\!\left(l\!+\!\!1\!\right)\!\right)\!\!\left(m_{2}^{2}c^{4}{a}^{2}\!+\!l\!\left(l\!+\!\!1\!\right)\!c^{2}{\hbar
}^{\!2}\right)\!}\right]^{\!\!{1\over 2}}
\eea
\noindent
The eigenenergies of a heteronuclear rotor are explicitly obtained from the above equation

\noindent
{\large \bf Non-relativistic limit}

\indent
To arrive at the non-relativistic limit of Eqn.(24), we first square the expression and then use the Taylor expansion 
for the term within the square root. This leads to 
\bea
W^{2} \ = \ \epsilon ^{2}\left[1+{4\over a^{2}}{l\left(l+1\right)c^{2}\hbar ^{2}\over \epsilon
^{2}}\ +{\left(m_{1}c^{2}-m_{2}c^{2}\right)^{2}\ l\left(l+1\right)c^{2}\hbar
^{2}\over \epsilon ^{2}\left\{
m_{1}c^{2}m_{2}c^{2}{a}^{2}+l\left(l+1\right)c^{2}\hbar ^{2}\right\} }\right]
\eea
where $\epsilon \ = \ m_{1}c^{2}+m_{2}c^{2}$ is the total rest energy of the system.

\indent
The relativistic correction is made evident by the binomial expansion of Eqn.(25) which gives 
\bea
W\approx \ \epsilon \ + \ {2l\left(l+1\right)c^{2}\hbar ^{2}\over \mathrel {a^{2}\epsilon
}}\ +{1\over
2}\ {\left(m_{1}c^{2}-m_{2}c^{2}\right)^{2}\ l\left(l+1\right)c^{2}\hbar
^{2}\over \epsilon \left(m_{1}c^{2}m_{2}c^{2}{a}^{2}+l\left(l+1\right)c^{2}\hbar
^{2}\right)}
\eea
\indent
While the first term corresponds to the rest energy of the system, the second refers to the NR term and the third gives 
first relativistic correction.

\indent
It is straightforward to check that 
\bea
E_{Nr}={2l\left(l+1\right)\hbar ^{2}\over I}\ {\mu \over M}
\eea
where $\mu =${\Large ${m_{1}m_{2}\over m_{1}+m_{2}}$}, $I=\mu a^{2}$ and $M=m_{1}+m_{2}$ as before.

\indent
In the case of equal masses, we recover the well-known expression for the energies of the non-relativistic rigid rotator.

\newpage
\noindent
{\large \bf Relativistic Rotational Coefficient}

\indent
In spectroscopic calculations, the rotational constant, B, is an important parameter, 
characteristic of the molecule$^{5}$. 
 It is seen that the energy spectrum of a non-relativistic rigid rotator consists of a series of 
equidistant lines, the first one lying at 2B and the separation of successive lines also being 2B.

\indent
Eqn.(26) may be written as
\bea
W_{KG}=\ \epsilon +l\left(l+1\right)hc\left[{h\over 4\pi ^{2}Ic}\ {2\mu \over
M}\left\{ 1+{\left(m_{1}-m_{2}\right)^{2}a^{2}\over
4\left(m_{1}m_{2}c^{2}{a}^{2}+l\left(l+1\right)\hbar ^{2}\right)}\right\}
\right]
\eea
\indent
Apparently, the second term within the brackets may rightly be called the `energy correction term' resulting due to unequal 
masses. This term highlights the effect of going from homonuclear to heteronuclear species.

\vspace{-0.1in}
\indent
The above equation may be put in a compact form as 
\vspace{-0.1in}
\bea
W_{KG}=\ \epsilon + \ l \left(l+1\right)hc\ B_{Rel}
\eea
where 
\bea
B_{Rel}={h\over 4\pi ^{2}Ic}\ {2\mu \over M}\left\{
1+{\left(m_{1}-m_{2}\right)^{2}a^{2}\over
4\left(m_{1}m_{2}{{c}^{2}a}^{2}+l(l+1)\hbar ^{2}\right)}\right\} 
\eea
$B_{Rel}$ may be called the $l$-dependent Relativistic Rotational Coefficient. In the case of equal masses, 
\vspace{-0.1in}
\bea
B_{Rel}={h\over 8\pi ^{2}Ic}\ ,
\eea
which is in agreement with the well-known rotational constant, B of non-relativistic  theory. Thus Eqn. (31) for 
homonuclear molecules is only a special case of the more general equation (Eqn.30) applicable to heteronuclear molecules. 

\noindent
{\large \bf Energy Spectrum}

\indent
Starting from Eqn.(25) and carrying out the Taylor expansion upto the first order, one can readily obtain 
\vspace{-0.1in}
\bea
W_{KG}=\ \epsilon
+ \ l\left(l+1\right)hcB_{Rel}-{l^{2}\left(l+1\right)^{2}h^{2}c^{2}B_{Rel}^{2}\over
2\epsilon }
\eea
which may be written as 
\vspace{-0.1in}
\bea
W_{l}=\ \epsilon
+ \ l\left(l+1\right)hc\left(B+B_{l}\right)-{l^{2}\left(l+1\right)^{2}h^{2}{c}^{2}\over
2\epsilon }\left(B+B_{l}\right)^{2}  
\eea
where 
\vspace{-0.1in}
\bea
B={h\over 4\pi ^{2}Ic}\ {2\mu \over M}
\eea
and 
\vspace{-0.1in}
\bea
B_{l}=B{\left(m_{1}c^{2}-m_{2}c^{2}\right)^{2}a^{2}\over
4\left(m_{1}c^{2}m_{2}c^{2}a^{2}+l\left(l+1\right)c^{2}\hbar ^{2}\right)}.
\eea
\indent
Essentially, for a rotor having unequal masses, each level is shifted differently. The rotational constant is trivially 
modified and $B_{l}$ may be interpreted as the relativistic correction term to the rotational constant, for a given transition. 

\vspace{-0.1in}
\indent
Eqn.(33) may also be written as 
\bea
W_{l}=W_{0}\>+\>l\left(l+1\right)hc\ B_{l}\,-\,{l^{2}\left(l+1\right)^{2}h^{2}c^{2}\over
2\epsilon }\left(B_{l}^{2}+2BB_{l}\right)
\eea
where 
\bea
W_{0}\,=\,\epsilon
+\ \!l\left(l+1\right)hc \,B \,\,- \,\,{l^{2}\left(l+1\right)^{2}h^{2}c^{2}{B}^{2}\over2\epsilon}
\eea
\indent
It is interesting to note that for equal masses, $B_{l}=0$ and $W_{l}=W_{0}$, the emperical formula obtained by
 Rao et.al.

\indent
Introducing the wave number, $\bar \nu _{l}=${\Large ${W_{l+1}-W_{l}\over hc}$}, we obtain for the transition $l+1\rightarrow l,$ 
\bea
\bar \nu _{l}=T_{1}+T_{2}+T_{3}+T_{4}+T_{5}
\eea  
where

\hspace{1.8in}$T_{1}=2\left(l+1\right)B$

\hspace{1.8in}$T_{2}=-${\Large ${2hc\over \epsilon }$}$B^{2}\left(l+1\right)^{3}$

\hspace{1.8in}$T_{3}=\left(l+1\right)\left\{ \left(l+2\right)B_{l+1}-lB_{l}\right\} $

\hspace{1.8in}$T_{4}=-${\Large ${\left(l+1\right)^{2}hc\over 2\epsilon }$}$\left\{
\left(l+2\right)^{2}B_{l+1}^{2}-l^{2}B_{l}^{2}\right\} $

\hspace{1.8in}$T_{5}=-${\Large ${\left(l+1\right)^{2}hc\over 2\epsilon }$}$2B\left\{
\left(l+2\right)^{2}B_{l+1}-l^{2}B_{l}\right\} $

\indent
A more rigorous calculation shows that the terms $T_{3},T_{4}$ and $T_{5}$ are mass-dependent and are explicitly given by 
\bea
T_{3}={B\left(l+1\right)m_{1}c^{2}m_{2}c^{2}\left(m_{1}c^{2}-m_{2}c^{2}\right)^{2}a^{4}\over
2\left(m_{1}c^{2}m_{2}c^{2}a^{2}+\left(l+1\right)\left(l+2\right)c^{2}\hbar
^{2}\right)\left(m_{1}c^{2}m_{2}c^{2}a^{2}+l\left(l+1\right)c^{2}\hbar
^{2}\right)} \ \ \ \ \ \ \ \
\eea
\bea
T_{4}=\!{B^{2}hc\left(l\!+\!1\right)^{2}\!\left(l\!+\!2\right)\!m_{1}c^{2}m_{2}c^{2}\!\left(m_{1}c^{2}\!-\!m_{2}c^{2}\right)^{2}\!a^{6}\!\left[{\!m}_{1}c^{2}m_{2}c^{2}a^{2}\!-\!2l\left(l+1\right)\!c^{2}\hbar
^{2}\right]\over 16\!\!\!\ \epsilon
\left(m_{1}c^{2}m_{2}c^{2}a^{2}+\!\left(l+\!1\right)\left(l+\!2\right)c^{2}\hbar
^{2}\right)^{2}\left(m_{1}c^{2}m_{2}c^{2}a^{2}+l\left(l+\!1\right)c^{2}\hbar
^{2}\right)^{\!2}}
\eea
\bea
T_{5}={B^{2}hc\left(l+2\right)^{2}\left(m_{1}c^{2}-m_{2}c^{2}\right)^{2}a^{2}\left\{
2l\left(l+1\right)c^{2}\hbar
^{2}+\left(l+2\right)m_{1}c^{2}m_{2}c^{2}a^{2}\right\} \over 2\ \epsilon
\left(m_{1}c^{2}m_{2}c^{2}a^{2}+\left(l+1\right)\left(l+2\right)c^{2}\hbar
^{2}\right)^{\!}\left(m_{1}c^{2}m_{2}c^{2}a^{2}+l\left(l+1\right)c^{2}\hbar
^{2}\right)} \ \ \
\eea

\indent
It is straightforward to check that $T_{3},T_{4}$ and $T_{5}$ vanish when the masses are equal. While the first two terms of 
Eqn.(38) correspond to the usual expression for the wave number of the spectral line, the last three terms arising mainly 
due to unequal masses do contribute to the small relativistic correction in the wave number.

\indent
Thus the application of relativistic quantum mechanics to the investigation of the spectra of heteronuclear linear diatomic 
molecules using the KG equation entails great mathematical complexities in all. 

\noindent
{\large \bf First Rotational Line}

\indent
In the spirit of non-relativistic approximation, we retain $T_{1}$ and $T_{3}$ and neglect all other terms in higher powers 
of $B,$ to obtain 
\vspace{-0.1in}
\bea
\bar \nu_{2}=2\left(l+1\right)B+\left(l+1\right)\left\{
\left(l+2\right)B_{l+1}-lB_{l}\right\} .
\eea 
\noindent
The first rotational line occurs at $l=0,$ leading to 
\vspace{-0.1in} 
\bea
\bar \nu _{0}\,=\,2B+2B_{1}.
\eea
It is trivial to check that
\vspace{-0.1in} 
\bea
\bar \nu _{0}\,=\,{B\over 2}\left[M^{2}{c}^{2}{a}^{2}\ +\ 8{\hbar }^{2}\over \mu
M{c}^{2}{a}^{2}\ +\ 2{\hbar }^{2}\right]
\eea
where $m_{1}m_{2}=\mu M$ is implied. In units of Compton wavelength expressing $\tilde a=${\Large ${a\over \hbar /Mc}$} and 
$\tilde a_{0}=${\Large ${a\over \hbar /\mu c}$}, we obtain 
\vspace{-0.1in} 
\bea
\bar \nu _{0}\,=\,{B\over 2}\left[\tilde a^{2}+8\over \tilde a\tilde
a_{0}+2\right].
\eea
\indent
For equal masses, $\bar \nu_{0}=2B,$ by virtue of Eqn.(42). It is thus seen that the allowed transitions 
for homonuclear diatomic molecules are regularly spaced at an interval of 2B. However,
for a heteronuclear KG rotor, the lines are no longer exactly equidistant. Apparently, the 
measurement and identification of one of the spectral lines allows us to calculate the 
moment of inertia and then the bondlength of the molecule.

{\section{\large\bf Results and Discussion}}

 A rigid rotator approximates a diatomic molecule when vibrational motion is ignored. In 
literature, the energy specturm of a rigid rotator is investigated non-relativistically 
using the Schrodinger equation. In the present article, we have introduced the
notion of the `relativistic rotator' and obtained the eigen energies relativistically. 
To investigate the energy spectrum of spin zero homonuclear and heteronuclear systems, we 
solve the Klein-Gordon equation in the centre of mass frame. Step by step
elimination of the square root of an operator results in the fourth order equation in $W,$ 
from which the eigen-energies are extracted. This novel approach can be  extended to spin 
half systems, as well.

\indent
Apparently, it is seen that the eigenenergies of heteronuclear systems reduce to those of 
homonuclear species when $m_{1}=m_{2}$, and have an approapriate non-relativistic limit. It 
is seen that the level spacing which is equidistant for
a non-relativistic rotator is not so for a relativistic rotor. Interestingly, 
the first rotational line  of a non-relativistic rotator and that of a Klein-Gordon 
rotor occurs at 2B. While it is known that the
pure rotation spectrum of a diatomic molecule consists of equispaced lines, that of a
 relativistic rotator is not so.

\indent
 It has been established that the rotational constant of homonuclear and heteronuclear 
diatomic molecules depend on the electronegativity of the constituent atoms $^{6}$. 
 Relativistic studies, apart from having immense pedagogical value are also of practical
use in the light of ever increasing accuracies of modern experimental methods. 

\indent
{\large\bf Acknowledgements}

    One of the authors(N. A. Rao) expresses her thanks  to Ms. M. V. Jayanthi, IAS, Commissioner for Collegiate Education in Karnataka  
for her endearing encouragement and support. Thanks are extended to Dr. K. N. Sreenivasa Rao, 
whose initial work on homonuclear rotor was a source of inspiration to the present work.

\newpage
\baselineskip 0.2in

{\large \bf References}
\begin{enumerate}

\item Herbert Goldstein, {\it Classical Mechanics}, Addison - Wesley Publishing Co. (1969).

\item Sauter F Ibid 69 (1931) 742.

\item K.N. Srinivasa Rao and D.Saroja, {\it The rigid rotator according to relativistic quantum mechanics}, Proc.Indian 
Academy of Sciences, Vol. LX1 No,{\bf 6} Sec. A (1965) 398 - 406.

\item Stephen Wolfram {\it The Mathematica Book}, Wolfram Media, 3rd edn. Cambridge University
Press, 1996.

\item G.Herzberg, {\it Molecular Spectra and Molecular Structure}-Vol, I Spectra of diatomic molecules - $2^{nd}$ Edition, 
D.Van Nostrand Co. Inc., New York, (1950), Chapter III.

\item C. Mande and V.G. Asolkar, {\it Dependence of rotational constants of homonuclear and intragroup diatomic molecules on 
electronegativity}, Proc. Nat. Acad. Sci. India {\bf 69(A)} II (1999) 237 - 245.
\end{enumerate}

\end{document}